\documentclass{epl}
\usepackage{amsmath,amssymb}
\usepackage{bm}
\usepackage{epsfig}

\newcommand{\llangle}{\langle\kern-.25em\langle}
\newcommand{\rrangle}{\rangle\kern-.25em\rangle}

\newcommand{\bk}{{\bf k}}
\newcommand{\bq}{{\bf q}}

\newcommand{\ve}{{\vec e}}

\newcommand{\bfk}{{\bf k}}
\newcommand{\bfr}{{\bf r}}

\newcommand{\zit}[1]{\cite{#1}}

\newcommand{\vn}{{\vec n}}
\newcommand{\vnl}{\vn}
\newcommand{\vnr}{\vn'}

\newcommand{\AAl}{a}
\newcommand{\AAlc}{\bar a}
\newcommand{\AAr}{a'}

\newcommand{\DA}{\Delta a}
\newcommand{\DAc}{\Delta \bar a}

\newcommand{\BBl}{b}
\newcommand{\BBlc}{\bar b}
\newcommand{\BBr}{b'}

\newcommand{\DB}{\Delta b}
\newcommand{\DBc}{\Delta \bar b}

\newcommand{\Pl}{\phi}
\newcommand{\Plc}{\bar \phi}
\newcommand{\Prx}{\phi'}
\newcommand{\Prc}{\bar \phi'}
\newcommand{\Dt}{\Delta \tau}

\newcommand{\cS}{{\mathcal S}}

\newcommand{\Gammap}{\Gamma^+}
\newcommand{\Gammam}{\Gamma^-}

\newcommand{\Jm}{J^-}
\newcommand{\Jp}{J^+}

\newcommand{\ud}{\textrm{d}}

\title{Frustrated Heisenberg antiferromagnets: fluctuation induced first order vs deconfined quantum criticality}
\author{F. Kr\"uger\inst{1} \and S. Scheidl\inst{2}}
\institute{
  \inst{1} Instituut-Lorentz, Universiteit Leiden, P.O. Box
9506, 2300 RA Leiden, The Netherlands\\
  \inst{2} Institut f\"ur Theoretische Physik, Universit\"at zu
K\"oln, Z\"ulpicher Str. 77, D-50937 K\"oln, Germany
}

\begin{document}

\maketitle

\begin{abstract}
Recently it was argued that quantum phase transitions can be radically
  different from classical phase transitions with as a highlight the 'deconfined 
  critical points' exhibiting fractionalization of quantum numbers due to Berry phase 
  effects. Such transitions are supposed to occur in frustrated ('$J_1$-$J_2$') quantum 
  magnets. We have developed a novel renormalization approach for such systems which is 
  fully respecting the underlying lattice structure. According to our findings, another
  profound phenomenon is around the corner: a fluctuation  induced (order-out-of-disorder)
  first order transition. This has to occur for large spin and we conjecture that it is
  responsible for the weakly first order behavior recently observed in numerical simulations 
  for frustrated $S=1/2$ systems. 
\end{abstract}

\title{Frustrated Heisenberg antiferromagnets...}

\section{Introduction}
Frustrated magnets exhibit quantum phase transitions of a rich variety
which is subject of intense current research \zit{Sachdev99}.  Novel
scenarios for phase transitions beyond the Landau-Ginzburg-Wilson (LGW)
paradigm have been suggested \zit{Senthil+04a,Senthil+04b} joggling fundamental
concepts. The Heisenberg model on a square lattice with
antiferromagnetic couplings $J_1$ and $J_2$ between nearest and
next-nearest neighbors serves as a prototype for studying magnetic
quantum-phase transitions (see, e.g., \zit{Sushkov+01} and references
therein).  From the classical limit one expects that two different
magnetic orders can exist: the N\'eel phase with ordering wave vector
$(\pi,\pi)$ is favorable for $\alpha \equiv J_2/J_1 <1/2$ and columnar
order with $(0,\pi)$ for $\alpha > 1/2$.

Quantum fluctuations certainly may induce a paramagnetic (PM) phase
which is naturally expected near $\alpha \approx 1/2$ where both
orders compete \zit{Chandra+88}.  Remarkably, additional orders may
appear in the N\'eel phase as well as in the PM phase when translation
symmetry is broken by an additional spin dimerization \zit{Sushkov+01}.  
The existence of such enhanced order crucially depends on the spin value $S$. 
This becomes most apparent when the spin system is represented by a
nonlinear-sigma model. Topological excitations associated with Berry phases 
can give rise to ground-state degeneracies corresponding to a translation-symmetry 
breaking by dimerization and formation of valence-bond solid (VBS) phases 
\zit{Haldane88,Fradkin+88}.

The N\'eel-VBS transition has been argued to be in a novel quantum criticality
class that does not fit in the standard LGW paradigm. Intriguing 
data on this transition was obtained in simulations of the $S=1/2$ quantum
XY model frustrated by ring exchange \zit{Sandvik+02,Melko+04}. 
The transition was interpreted as a second-order one; this possibility 
was predicted by the theory of the deconfined critical point and suggested
to be generic for a variety of experimentally relevant two-dimensional 
antiferromagnets \zit{Senthil+04b}. However, in a more careful finite-size 
analysis of the XY case it was demonstrated that the N\'eel-VBS point
represents an anomalously weak first-order transition \zit{Kuklov+04}. 

Field-theoretical approaches of various kinds have been developed,
based on the $1/S$ \zit{Chakravarty+89} or the $1/N$ expansion
\zit{Read+89,Read+90}.  The latter approach is able to capture some of the
essence of the topological aspects on a mean-field level.  However, it
is the former approach, elaborated to a renormalization-group
analysis, which predominantly has served as basis for a comparison of
critical aspects between theory and experiment (see e.g.
\zit{Hasenfratz00}).

Nevertheless, this approach so far has suffered from two intrinsic
shortcomings: i) Spin-wave interactions are the physical mechanism
underlying the renormalization flow.  On one-loop level, the flow
equations describe corrections to the physical parameters of relative
order $1/S$.  These corrections have been worked out using a continuum
version of the nonlinear $\sigma$ model (CNL$\sigma$M) as a starting
point \zit{Chakravarty+89}.  However, for the original lattice model,
this is not systematic, since corrections of the same order are
dropped under the naive coarse graining of the lattice model onto the CNL$\sigma$M.
As a consequence, important effects such as a renormalization of the
spin-wave velocity and the frustration $\alpha$ are missed.  ii) 
Similarly, the outer large momentum region of the magnetic Brillouin zone (BZ) 
is only crudely treated. This entails a significant 
uncertainty in the computation of the phase diagram.

In this Letter, we lift these shortcomings by developing a
renormalization analysis which fully accounts for the lattice
structure. It combines the systematic treatment of all corrections in
order $1/S$ on the level of the conventional first-order spin-wave
theory (SWT) with the merits of a renormalization approach, which goes
beyond any finite order in $1/S$ by an infinite iteration of
differential steps, successively eliminating the spin-wave modes
of highest energy.  As a result, we obtain an improved description of
the phase transitions. In particular, the possibility of a fluctuation 
induced first-order transition which is not accessible on the level of 
the CNL$\sigma$M is included in a natural way.

\section{Spin-coherent state representation}
The key to what follows is a novel kind of path integral quantization, which 
makes it possible for us to treat the effects of umklapp on an equal
footing with the spi-wave interactions.
To be specific, we stick to the aforementioned $J_1$-$J_2$ Heisenberg
model on a square lattice.  We address the stability of the N\'eel
phase against quantum fluctuations controlled by $S$ and $\alpha$.
Fluctuations are treated in a coherent spin state path-integral
representation of the model, where a spin state corresponds to a unit
vector $\vn$.  In the absence of fluctuations, spins would assume the
states $|\vn_{A/B}\rangle = |\pm \ve_z\rangle = |S,\pm S\rangle$ on
the two sublattice A and B.  From the standard Trotter formula emerges
an imaginary time $\tau$ (discretized in intervals of duration $\Dt$)
and an action of the form \zit{Sachdev99}
\begin{eqnarray}
  \cS = - \sum_\tau \ln \langle \{\vn\} | \{\vn'\}\rangle
  + \sum_\tau \Dt  \frac { {\langle \{\vn\} | \hat H |\{\vn'\}\rangle}}
   {\langle \{\vn\} | \{\vn'\}\rangle}
\label{action}
\end{eqnarray}
taking $\vn$ at time $\tau$ and $\vn'$ at $\tau-\Dt$.  For weak
quantum fluctuations, the components of $n_{x,y}$ perpendicular to
the magnetization axis are small and may be considered as expansion
parameters (as in \zit{Chakravarty+89}).  However, attempting to
directly apply this expansion to the lattice model we encountered time
ordering difficulties in the action \zit{KS-unpub}.

Instead, we start form a stereographic parametrization of coherent
states, on sublattice A, $|\vn \rangle = (1+\AAlc\AAl/2S)^{-S}
\exp(\AAl \hat S_-/\sqrt{2S}) |S,S\rangle$ where $a$ is the
stereographic projection of $\vn$ from the unit sphere onto the
complex plane, $\AAl = \tan(\theta /2) \exp(i \phi)$ with the standard
spherical angles $\theta$ and $\phi$.  The action can be expressed in
terms of the stereographic coordinates using the matrix
elements\zit{Radcliffe71}
\begin{eqnarray}
  \frac {\langle \vnl | \hat S_z| \vnr \rangle}
  {\langle \vnl | \vnr \rangle}
  = S \frac{1-\AAlc \AAr/2S}{1+\AAlc \AAr/2S}, \quad
  \frac {\langle \vnl | \hat S_+| \vnr \rangle}
  {\langle \vnl | \vnr \rangle}
  = \frac{\sqrt{2S} \ \AAr}{1+\AAlc\AAr/2S}
\end{eqnarray}
Here $\bar a$ is the complex conjugate of $a$. The expressions
for the coordinate $b$ on sublattice B are given by the correspondences
$\hat S_z\to -\hat S_z$ and $\hat S_\pm\equiv\hat S_x\pm\hat S_y\to\hat S_\mp$
for $a\to b$.

The explicit expression of the action in terms of $a$ and $b$ is too
lengthy to be given here.  To leading order in $1/S$, the fluctuations
are controlled by the bilinear part $\cS_0$ of the action that
represents free magnons.  We also retain the quartic contribution
$\cS_{\rm int}$ to the action, which represents magnon interactions
and contain the renormalization of single-magnon parameters of
relative order $1/S$.  Higher order contributions to the action are
neglected.  Terms from the functional Jacobian are also negligible on
this level.  The single-magnon contribution to the action can be
parameterized in the form
\begin{eqnarray}
  \cS_0 &=& \sum_\tau \int_\bk \{ \frac 1{2g}
  [\AAlc_\bk \DA_\bk - \DAc_\bk \AAr_\bk
  + \BBlc_\bk \DB_\bk - \DBc_\bk \BBr_\bk]
  \nonumber \\ &&
  +\Dt S [\Gammap_\bk  (\AAlc_\bk \AAr_\bk +\BBlc_\bk \BBr_\bk)
  + \Gammam_\bk  (\AAlc_\bk \BBlc_\bk + \AAr_\bk \BBr_\bk)]
  \},\quad
\end{eqnarray}
using the Fourier transform $ \AAl(\bfr)= \int_\bk e^{i \bk \cdot \bfr
} \AAl_\bk$, the intgeral $\int_\bk = 2 \int \frac{d^2k}{(2\pi)^2}$
over the magnetic BZ $|k_x|+|k_y|\leq \pi$, and the exchange couplings
$\Gammap_\bk \equiv \Jp_\bk - \Jp_0 + \Jm_0$, $\Gammam_\bk \equiv
\Jm_\bk \equiv 2 J_1 [\cos(k_x)+\cos(k_y)]$ and
$\Jp_\bk=2J_2[\cos(k_x+k_y)+\cos(k_x-k_y)]$. For simplicity the
lattice constant is considered as unit length.  The dimensionless
parameter $g$ represents the strength of quantum fluctuations.  It
assumes the value $g=1$ in the unrenormalized model and turns out to
increase under renormalization.

Diagonalizing this bilinear action, one easily obtains the magnon
dispersion $E_\bk = S g |\Gammap_\bk|\\ \sqrt{1-\gamma_\bk^2}$, where
$\gamma_\bk=\Gammam_\bk/\Gammap_\bk$. For $\alpha \leq 1/2$, the
low-energy spin-wave excitations are characterized by an isotropic
dispersion $E(\bk) = c |\bk| + O(k^2)$ with a spin-wave velocity $c =
\sqrt 8 g S J_1 \sqrt{1-2\alpha}$.  Likewise, the exchange couplings
generate a spin stiffness $\rho = S^2 J_1 (1-2\alpha)$ for low-energy
modes.  One also obtains the propagators $\langle \Plc_\bk \Prx_\bk
\rangle = g G_\bk$ and $ \langle \Plc_\bk \Pl_\bk \rangle = \langle
\Prc_\bk \Pl_\bk \rangle = g (G_\bk+1)$ for fields from the same
sublattice ($\phi=a,b$), and $ \langle \AAl_\bk \BBl_\bk \rangle=
\langle \AAlc_\bk \BBlc_\bk \rangle = - g F_\bk$ for fields from
different sublattices. In the latter case the correlators are
unchanged by a replacement $\bar \phi \to \bar \phi'$.  We have
defined $G_\bk = (n_\bk+1/2) (1-\gamma_\bk^2)^{-1/2} - 1/2$ and $F_\bk
= (n_\bk+1/2) \gamma_\bk (1-\gamma_\bk^2)^{-1/2}$ where $n_\bk =
(e^{\beta E_\bk}-1)^{-1}$ is the bosonic occupation number.  For
strong frustration ($\alpha>1/2$) the stiffness becomes negative and
the spin-wave velocity is ill defined due to the presence of unstable
modes in the center of the BZ (see Fig. \ref{fig.rbz}).

\section{Renormalization approach}
\begin{figure}[htbp]
  \centering
  \epsfig{file=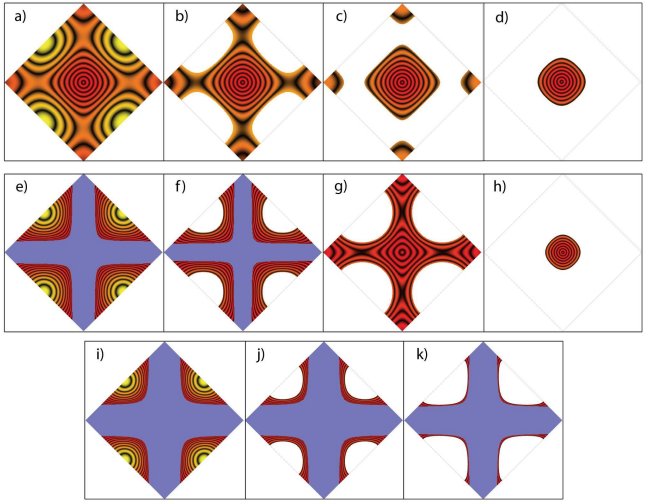, width = 0.6\linewidth}
    \caption{
      Evolution of modes under coarse graining. Each panel corresponds
      to the area $|k_x| \leq \pi$, $|k_y| \leq \pi$.  Red color
      represents small, yellow high positive energy. Black lines are
      lines of constant energy, blue areas represent unstable modes.
      Panels (a)-(d): In the N\'eel phase the RBZ may become
      disconnected first, then always shrinks to a sphere (here
      $S=1/2$, $\alpha=0.3$, $l=0, 0.25, 0.5, 1.0$).  Panels (e)-(h):
      In the N\'eel phase for strong frustration $\alpha>1/2$,
      initially unstable modes (blue area) are renormalized to stable
      modes and the RBZ eventually also shrinks to a sphere (here
      $S=1/2$, $\alpha=0.55$, $l=0, 0.11, 0.29, 1.30$).  Panels
      (i)-(k): In the columnar phase, after the elimination of all
      stable modes, an area of unstable modes survives (here $S=2$,
      $\alpha=0.6$, $l=0, 0.10, 0.20$).  }
\label{fig.rbz}
\end{figure}

Starting from this action with bilinear and quartic terms, we
implement a renormalization procedure as follows.  In successive
steps, the modes of highest energy (an infinitesimal fraction of all
modes) are integrated out. This decimation of modes yields an
effective theory for the remaining modes and gives rise to
differential flow equations.  As flow parameter we choose $l = \frac
12 \ln (A_{\rm BZ} / A_{\rm RBZ})$, where $A_{\rm BZ}=2 \pi^2$ is the
area of the original BZ, and $A_{\rm RBZ}$ is the area of the residual
BZ (RBZ) populated by the remaining modes. On large length scales in
the N\'eel phase, the RBZ becomes circular and $l$ is the usual
logarithmic length scale.  The evolution of the RBZ and the
single-magnon dispersion is illustrated in Fig.~\ref{fig.rbz} for
various parameters.

To one-loop order, corresponding to a systematical calculation of
corrections in order $1/S$, the renormalization effects due to
spin-wave interactions can be captured by a flow of the single-magnon
parameters.  The resulting flow equations are given by

\begin{eqnarray}
  \ud g^{-1} &=& - \frac 1S  \ud G^0,
  \\
  \ud J_1 &=& \frac g S J_1 ( \ud F^- - 2  \ud G^0),
  \\
  \ud J_2 &=& \frac g S J_2 ( \ud G^+ - 2  \ud G^0), 
\end{eqnarray}

where the integrals over the differential fraction $\partial$ of modes
of highest energy are defined as $ \ud G^0 = \int_\bq^\partial G_\bq$,
$ \ud G^+ = \int_\bq^\partial (J^+_\bq/J^+_0) G_\bq$, and $ \ud F^-=
\int_\bq^\partial (J^-_\bq/J^-_0) F_\bq$.

Since the BZ does not remain self-similar under mode decimation, we
omit the usual rescaling of length and time which is necessary only
for the identification of fixed points under a renormalization-group
flow.  However, dropping this rescaling does not lead to a loss of
information. Then, each fixed point represents an antiferromagnetically
ordered state. The quantum-disordered phase and the transition into it
show up as run-away flow.

\section{Results and discussion}
Because of the changing geometry of the RBZ and the incorporation of
the full single-magnon dispersion in our renormalization approach, the
flow equations can be integrated  only numerically.  Here, we focus
on $T=0$, although the flow equations are valid also for $T >0$.  The
flow of parameters is characterized by the following tendencies.  Both
exchange couplings $J_{1,2}$ as well as $1/g$ always shrink.  These
fundamental parameters flow in such a way that $\alpha$ always
decreases, $c^2$ increases (initially it is negative for
$\alpha>1/2$), while $\rho$ may increase for small $l$ until it
decreases for sufficiently large $l$.

\begin{figure}[htbp]
  \centering
  \epsfig{file=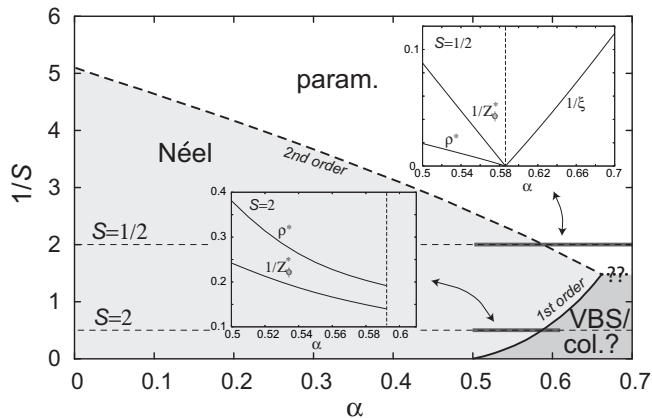, width=0.6 \linewidth}
  \caption{Phase diagram showing the 2nd order transition line 
    between the N\'eel ordered and the PM phase (dashed line) and the
    1st order boundary between the N\'eel and the columnar phase
    (solid line).  The border between the PM and the
    columnar phase (dotted line) is just guesswork and cannot be
    calculated within our approach. Insets: Nature of the phase
    transitions. For $S=1/2$ the renormalized spin stiffness $\rho^*$
    vanishes and the strength of the quantum fluctuations $Z_\phi^*$
    diverges at the phase boundary corresponding to a 2nd order
    transition. Close to the the N\'eel phase the correlation length
    diverges like $\xi\sim (\alpha-\alpha_c)^{-1}$. At the 1st order
    transition ($S=2$) $\rho^*$ and $Z_\phi^*$ remain finite. }
  \label{fig.phase}
\end{figure}

The nature of magnetic order can be identified from the flow behavior.
Three possibilities are observed. (i) The RBZ shrinks to a circle of
decreasing radius $\propto e^{-l}$ [see Fig.~\ref{fig.rbz} panels
(a)-(d) and panels (e)-(h)] while $J_{1,2}$ and $g$ (as well as the
derived quantities $c$, $\rho$, $\alpha$) converge to positive values.
Then N\'eel order is present, characterized by these renormalized
low-energy parameters. (ii) At some finite $l^*$, fluctuations become
so strong that $g$ diverges and $J_{1,2}$ vanish.  This indicates the
loss of magnetic order due to overwhelming quantum fluctuations.
Close to the transition to the N\'eel phase, the magnetic correlation
length -- which can be identified with $\xi = e^{l^*}$ -- diverges
algebraically like $\xi\sim (\alpha-\alpha_c)^{-1}$. For $S=1/2$ this
asymptotic behavior is shown in one inset of Fig.~\ref{fig.phase}.
(iii) For strong frustration, it is also possible that a finite RBZ of
unstable modes remains after decimation of {\it all} stable modes (see
Fig.~\ref{fig.rbz} panels (i)-(k)).  This indicates that the
instability towards columnar order is effective for the renormalized
low-energy modes.

The region of stability of the N\'eel phase is illustrated by the
light grey region in Fig.~\ref{fig.phase}.  In the absence of
frustration, we find N\'eel order to be stable for $1/S \lesssim 5.09$
in remarkable agreement with conventional linear SWT \zit{Chandra+88}.
However, this is pure coincidence since in linear SWT the phase
boundary is determined by the vanishing of the local magnetization
calculated in order $S^0$, whereas here it is determined by the
divergence of $g$ due to spin-wave interactions treated in one-loop
order. While the phase boundary is located at academically small spin
values for small frustration, $S$ reaches physically meaningful values
at larger frustration where the discrepancies between SWT and our
renormalization approach become more pronounced. In linear SWT, the
phase boundary smoothly approaches $1/S \searrow 0$ for $\alpha \nearrow 0.5$,
whereas we find the N\'eel phase to reach up to $\alpha = 0.66$ for
$S= 0.68$. For spins smaller than this value, the N\'eel phase becomes
unstable towards a PM phase via a second-order transition, whereas it
becomes unstable for $S>0.68$ via a first order
transition. Since we deal with a discontinuous transition we can only 
speculate about the type of ordering in the adjacent phase (dark shaded 
region in Fig. ~\ref{fig.phase}). From the classical limit we expect columnar
order for very large $S$ whereas for intermediate spin also VBS order may be 
present.

In the region where the N\'eel phase reaches up to $\alpha>1/2$,
initially unstable modes are renormalized to stable ones by spin-wave
interactions. Simultaneously, $\alpha$ is renormalized to a value
$\alpha^*<1/2$ and the flow behavior (i) is realized.  In the columnar
phase, the flow of $\alpha$ saturates at a value $\alpha^*>1/2$ and
the flow behavior (iii) is observed.

Stability of N\'eel order for $\alpha>0.5$ so far has been found only
by a self consistently modified SWT \zit{Xu+90} and Schwinger-boson
mean-field theory (SBMFT) \zit{Mila+91}.  The overall shape of the
N\'eel phase of these approaches is consistent with our findings.
However, in modified SWT and SBMFT the first-order transition from
N\'eel to columnar order can be obtained only by a comparison of free
energies between the two phases.  In our theory, the transition
directly emerges from the analysis of spin-wave interactions in the
N\'eel phase.

\begin{figure}[htbp]
  \centering
  \epsfig{file=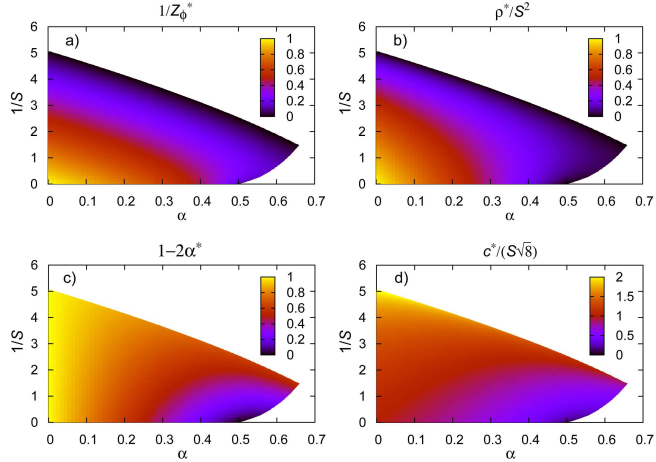, width = 0.6\linewidth}
   \caption{
     Values of the renormalized parameters within the N\'eel phase.
     Color corresponds to the strength of large-scale fluctuations (a),
     the renormalized spin stiffness (b), the frustration (c), and the
     spin-wave velocity (d).}
  \label{fig.ren}
\end{figure}

The nature of the transitions out of the N\'eel phase becomes apparent
from the behavior of the fluctuations on large length scales ($k \to
0$), where $\langle \bar \phi_\bfk \phi_\bfk' \rangle \simeq \frac
{Z_\phi}{\sqrt2 k }$ with $Z_\phi \equiv \frac {g}{\sqrt{1-2\alpha}}
\propto \frac c\rho$.  Approaching the transition into the PM phase,
the renormalized value $Z_\phi^*$ diverges and gives rise to a
divergent susceptibility (see Fig.~\ref{fig.ren}).  At the same time
the renormalized $\rho^*$ vanishes while $c^*$ remains finite.  The
continuous evolution of $Z_\phi^*$ and $\rho^*$ indicate the
second-order nature of the transition.  Approaching the transition
into the columnar phase, one observes a saturation of $Z_\phi^*$ and
$\rho^*$ at finite values as well as a discontinuous jump of $\alpha^*$
indicating a first-order transition.  Fig.~\ref{fig.ren} illustrates
the dependence of various renormalized quantities on $\alpha$ and
$1/S$ within the N\'eel phase. The insets of Fig.~2 show the evolution
of $\rho^*$, $Z_\phi^*$ and $\xi$ with higher resolution at the
transitions.

Confidence in the reliability of our findings is provided by
quantitative comparisons for specific parameter values.  Results for
$Z_c^* \equiv c^*/c$ exist from various approaches. For $S=1/2$ and
$\alpha=0$, first-order SWT yields a slight enhancement of spin-wave
velocity, $Z_c^{\rm SWT} = 1.158$.  We find an increased value $Z_c^*
= 1.20$, which is in agreement with Monte Carlo (MC) simulations (see
\cite{Manousakis91} and references therein).  As $1/S$ and/or $\alpha$
increases, the enhancement factor $Z_c^*$ also increases.  At the
phase boundary $1/S \approx 5.09$ for $\alpha=0$, the difference is
already more pronounced: $Z_c^{\rm SWT} = 1.40$ and $Z_c^* = 2.04$.
For $\alpha>0$, unfortunately, MC data for $Z_c$ are not available at
the transition out of the N\'eel phase, neither for $S=1/2$ nor for
larger physical values of $S$.

\section{Conclusion}
We have presented a novel renormalization approach for
frustrated quantum antiferromagnets which fully accounts for the
underlying lattice geometry and consistently captures the
renormalization of the single-magnon parameters by spin-wave
interactions all over the magnetic BZ. 

For the $J_1$-$J_2$ model 
we clearly demonstrated that for large spins and strong frustration, fluctuations 
on lattice and intermediate scales cause an instability of the N\'eel phase 
towards a first order transition. These effects are totally missed by any effective 
long-wavelength continuum theory 
obtained by a naive coarse graining of the lattice model.

We conjecture that these fluctuations which crucially depend on the underlying 
lattice geometry and the way of frustrating the N\'eel order are responsible 
for the weakly first order behavior observed recently in numerical simulations 
for frustrated $S=1/2$ systems \zit{Kuklov+04}.  

\acknowledgments
The authors benefited from stimulating discussions with J. Zaanen, J. Betouras and 
A. Sandvik and thank J. van Wezel for critical reading of the manuscript. 
This work was supported by Dutch Science Foundation NWO/FOM and by 
Deutsche Forschungsgemeinschaft SFB 608.

\end{document}